**Towards a criteria-based approach to selecting human-AI interaction mode**

Jessica Irons[1], Patrick Cooper[1], Melanie McGrath[1], Shahroz Tariq[1] and Andreas Duenser[1]

[1]Commonwealth Scientific & Industrial Research Organisation (CSIRO), Australia

Contact: Jessica Irons, jessica.irons@csiro.au

**Abstract**

Artificial intelligence (AI) tools are now prevalent in many knowledge work industries. As AI becomes more capable and interactive, there is a growing need for guidance on how to employ AI most effectively. The A$^2$C framework (Tariq, Chhetri, Nepal & Paris, 2024) distinguishes three decision-making modes for engaging AI: automation (AI completes a task, including decision/action), augmentation (AI supports human to decide) and collaboration (iterative interaction between human and AI). However, selecting the appropriate mode for a specific application is not always straightforward. The goal of the present study was to compile and trial a simple set of criteria to support recommendations about appropriate A$^2$C mode for a given application. Drawing on human factors and computer science literature, we identified key criteria related to elements of the task, impacts on worker and support needs. From these criteria we built a scoring rubric with recommendation for A$^2$C mode. As a preliminary test of this approach, we applied the criteria to cognitive task analysis (CTA) outputs from three tasks in the science domain – genome annotation, biological collections curation and protein crystallization– which provided insights into worker decision points, challenges and expert strategies. This paper describes the method for connecting CTA to A$^2$C, reflecting on the challenges and future directions.

keywords: *Artificial intelligence, Cognitive task analysis; Human-AI collaboration.*

**Towards a criteria-based approach to selecting human-AI interaction mode**

The last decade has seen immense growth in the use of AI-based tools for knowledge work, from domain-specific decision aids in domains such as healthcare to general purpose generative AI tools (e.g. ChatGPT) employed across a wide range of the service sector. Increasingly AI is incorporated into commercial software such as the Microsoft suite of tools, meaning more workers than ever are able to interact with the technology. As the availability of AI grows, so too does interest in human-AI collaboration. While the concept of collaborating with AI (and automation, autonomy and machines in general, henceforth referred to as AI) has been discussed in human factors and related domains for many years, the accessibility and interactivity of generative AI has popularised the idea beyond research and into the workplace. The term recognises that as we become more reliant on technology, and as technology becomes more intelligent, our interactions have the potential to become closer to those between human team members.

A major challenge in the successful realisation of human-AI collaboration in the work context is knowing when and how to employ it to the greatest effect. Not all tasks are suitable for end-to-end automation, nor do they always require a long dialogue with a large language model. Guidance on effective use is particularly lacking for generative AI and agentic systems (teams of coordinated generative AI). While controlled lab studies have demonstrated benefits on task efficiency in coding (Peng, Kalliamvakou, Cihon & Demirer, 2023) and writing (Noy & Zhang, 2023), productivity and economic gains at the organisational level do not always benefit (Simkute et al., 2024). A study of workers across ten work domains found that although a large proportion of the sample believed that generative AI tools would halve the time required for a given task, less than 50% intended to use it (Humlum & Vestergaard, 2024). Perceived efficiency gain does not appear to be not enough to translate into uptake.

Determining the role for AI a knowledge work task begins with a deeper understanding of the task context: the workflow, existing challenges and needs of the stakeholders (Isagah & Ben Dhaou, 2023; Salwei & Carayon, 2022). This usually involves engaging with experts to understand the work as performed in practice. Human factors and engineering specialists can apply formalised methods to produce a detailed analysis of the task that can inform software requirements and allocation of functions to humans and machines (Hoffman, 2008; Rasmussen, 1986; Roth & Woods, 1989; Roth, Sushereba, Militello, Diiulio & Ernst, 2019). This process can take significant time and expertise that small-scale design projects may be unable to support. There are few simple guidelines to help translate insights about a task into actionable decisions on how to leverage AI effectively. In the present study, we report on an attempt to build and trial a set of criteria to guide human-AI collaboration decisions, using the human-AI interaction mode framework A$^2$C (Tariq, Chhetri, Nepal & Paris, 2024).

**Human-AI interaction modes**

Research on different human-AI interaction modes has often been characterised in terms of "levels of automation": incremental stages varying the degree of human and machine input on a task (Endsley, 1987; Endsley & Kaber, 1999; Frohm, Lindström, Stahre & Winroth, 2008; Parasuraman, Sheridan & Wicken, 2000; Sheridan & Verplank, 1978; Van Wezel, Cegarra & Hoc, 2011). For example, Sheridan and Verplank (1978), and later Parasuraman et al., (2000), outlined ten levels of automation from manual to full automation (where the machine completes the entire task without human intervention). Intermediate steps varied according to the machine's actions (suggest, recommend or make decisions); how the machine informs the human; and the human's role (decide, approve or veto machine decisions). While extremely influential, levels of automation are contentious, in part because they do not speak to the nature of the interaction or teamwork between agents (Johnson et al., 2011).

The A$^2$C framework (Tariq et al., 2024) captures high-level modes of interaction to inform software design. The framework comprises three levels – automation, augmentation and collaboration. Automation refers to the case where a task is performed entirely by an automated system, with the option to inform the human of its decision and reasoning process. Augmentation occurs where an AI system provides information or recommendation to support a human task, with the human ultimately responsible for any decision-making (see also Raisch & Krakowski, 2020; Choudhary, Marchetti, Shrestha & Puranam 2023). Collaboration occurs when the human and machine agents engage in a sustained period of back-and-forth interaction that ultimately leads to a decision.

Collaboration can be distinguished from augmentation in two key ways. Firstly, rather than a one-way flow of information (from AI to human), collaboration involves reciprocal communication or commands between the human and AI (Schleiger, Mason, Naughtin, Reeson & Paris, 2024). Secondly, collaboration is more than simply a cyclical iteration of the same process. Rather, each collaborative iteration builds on the last. This requires both the AI and the human to adapt the functions they perform in response to the information received from their partner.

These three different AI modes require different forms of technology with wide-ranging implications for infrastructure, costs and labour. How then should organisations make decisions about the mode appropriate for a given task? Principles of function allocation emphasise considering the impact of technology change on human performance (e.g. Endsley 2017; Parasuraman et al., 2000; Parasuraman, 2000). Typically, this process will begin with a task analysis, a method for decomposing, understanding and representing a task in consultation with subject matter experts. Building on this, Roth et al. (2019) propose adding consideration of interdependencies between humans and technology and the implications of different types of work allocations. In the information management literature, researchers have advocated matching technology to appropriate problems by examining the degree to which state-of-the-art automation tools can meet task needs (Engel, Elshan, Ebel & Leimeister, 2024; Hofmann, Jöhnk, Protschky & Urbach, 2020). While these approaches are typically very detailed or require an understanding of the technology landscape, from this research, it seems feasible that a set of simple criteria to guide organisations in deciding AI mode could be derived.

**Aim of this study**

This study aimed to addressed three research questions:

> *RQ1: What are the criteria to guide selection of human-AI interaction mode?*
>
> *RQ2: Can we address these criteria for a given task using the outputs from a task analysis?*
>
> *RQ3: (How) can we then bring together these criteria to make a recommendation about interaction mode needed?*

The paper is structured as follows: first, we review the existing literature of task factors relevant to automation mode. Next, we report on an application of the criteria to cognitive task analysis outputs from real world case studies. The case studies were three tasks from different science domains: genome annotation, protein crystallisation and digitisation of biological collection. We present our AI mode recommendations for the three use cases and themes across the use cases. We then review the process of applying the criteria and discuss developments for future iterations of the rubric.

**Literature review: Criteria informing Human-AI mode**

We first conducted a literature review to identify factors associated with performance under different human-AI modes in the $A^2C$ framework. At this initial stage, we concentrated on automation of cognitive tasks, which form the major component in knowledge work domains (Pyöriä, 2005). We also focused the review on factors stemming from the current task context rather than a desired future state, in line with our goal to design a tool that could be applied early in the design process. Other factors concerning the feasibility of implementing different AI technologies (e.g. availability of training data) and those arising from consideration of the future state (e.g. the need for training, trust in AI) are therefore not captured fully here, but are still critical to consider in the broader design process.

The factors fell within three broad categories: elements of the task, impacts on workers, and the existing challenges and needs that drive stakeholders to consider using AI.

**Task elements**

*Task type*

Analysis of levels of automation have often been accompanied by a breakdown of the type(s) of task required of the human or machine agents, conceptualised from a cognitive perspective. We used the characterisation by Parasuraman et al. (2000) that describes four task components: information acquisition, information analysis, decision and action (see also Endsley & Kaber, 1999 for a similar framework). Type of task interacts with other factors in making predictions about when and how AI should be implemented (as discussed below). As a basic principle, evidence suggests automating information acquisition, analysis and action implementation produce more benefits, fewer issues and greater acceptance than automating decision-making (e.g. Duan, Edwards & Dwivedi, 2019; Endsley, 2017; Onnasch, Wickens, Li & Manzey, 2014)

*Task complexity*

Various taxonomies of task complexity have been proposed (see Liu & Li 2012). Here we adopt the framework by Wood (1986) that defines three sources of complexity. Component complexity concerns the number and variety of steps in a task and the distinct sources of information to be processed. Coordinative complexity refers to the relationships between parts of the task and interdependencies within them. High coordinative complexity often necessitates scheduling or resource distribution components within the task (Crowston & Bolici, 2019; Malone & Crowston, 1994). Lastly, dynamic complexity arises when changes in the world occur mid-task and impact on the task, which introduces a need to monitor and adapt to change.

Van Wezel et al. (2011) suggest the AI inclusion on complex tasks is beneficial. For information acquisition and analysis task in particular, AI decision support can reduce complexity and support more effective decision-making (Jarrahi, 2018; Salimzadeh, He & Gadiraju, 2023; Sitchenko & Coiera, 2003). However, efforts to map task complexity to automation suggest that the higher the level of complexity, the less suitable the task for full automation (e.g. Peng & Bhaskar, 2023, Salimzadeh et al., 2023). Evidence suggests automated systems are typically brittle and do not handle dynamic or highly interdependent scenarios well (Cummings & Bruni, 2009). The ability for a human to identify and correct AI errors become increasingly difficult as complexity increases (Van Wezel et al., 2011).

*Variability*

Related to complexity is the variability in how a task is performed. While some tasks are triggered under a small set of circumstances and conducted in a routinised manner, others respond to a much broader set of initial conditions and are performed in different ways. Routine tasks are typically much more amendable to automation, as the necessary conditions can be defined computationally. Evidence suggests automating such tasks improves performance (Goh et al., 2019; Liu & Li , 2012; Onnasch et al., 2014; Perrow, 1967). Auter, Levy & Murnane (2003) suggested that while machines are feasible substitutes for humans in routine tasks (i.e. full automation), they should complement humans in non-routine tasks (augmentation or collaboration).

Task variability may also be a key factor in determining when collaboration may add value over simple augmentation. Tariq et al. (2024) argued that collaboration may be most appropriate in novel circumstances, in which neither the human nor the AI have a ready solution. In such circumstances the AI could assist the human by gathering information and assisting problem solving.

*Uncertainty*

Uncertainty is defined from the perspective of the task performer as the perceived inability to make accurate predictions (Milliken, 1987). Two key sources of uncertainty are a lack of information about the current problem or the available actions, and a lack of understanding of the structure of the problem state and the impact of different actions (Crowston & Bolici, 2019; Lipshitz & Strauss, 1997).

The presence of uncertainty leads to slower and less accurate decision-making for both human and machines (Platt & Huettel, 2008). Nevertheless, humans can employ effective coping strategies to make decisions in the face of uncertainty, such as evaluating pros and cons and employing rules of thumb, previous experience and intuition (Klein, 2008; Lipschitz & Strauss, 1997). Task experts can efficiently arrive at a "good enough" solution that achieves a desired outcome. AI may play a critical role in reducing uncertainty by gathering additional information (Cummings & Bruni, 2009); however, the human capacity to make decisions in the face of uncertainty has led researchers to suggest that the greater the uncertainty, the more reason to keep the human in the loop (Choudhary et al., 2023; Cummings & Bruni, 2009; Jarrahi, 2018).

*Non-codified knowledge*

Some tasks rely heavily on information that is not codified, but exists as experiential, tacit or procedural knowledge of experts (Klein, 2008). This form of knowledge can be difficult for rules-based AI applications. While knowledge elicitation methods can help bring this information to light, often the rules for applying tacit knowledge are hard to define. It may also imply that the task is ambiguous or complex. Cummings (2014) suggested that tasks reliant on non-codified knowledge should retain human involvement, but that AI can assist through collecting and synthesising information.

**Impacts on workers**

*Maintaining situation awareness and skills*

Many tasks require a worker to build situation awareness, defined as the ability to perceive elements of the environment, understand their meaning and predict future changes (Endsley, 1995). Automating a task does not automatically remove the need for situation awareness; awareness built on one task may be required for subsequent tasks, or the human may be required to takeover if the AI fails. The impact of automation on situation awareness depends on task type: automating information acquisition and, in some cases, information analysis can improve situation awareness, while automating the decision reduces awareness (Endsley, 2017; Endsley & Kiris, 1995; Endsley & Kabar, 1999; although see Tatasciore, Bowden, Visser, Michailovs & Loft, 2020). If the human is brought into the process to correct the AI, having been excluded thus far, they may lack the necessary awareness to complete the task, known as an "out-of-the-loop error" (Endsley, 2017). Being actively involved in decision-making process through augmentation or collaboration can support effective recovery in the case of automation failure (Onnasch et al., 2014).

Related to situation awareness is maintenance of skills and knowledge required to complete a task. Full automation is not recommended if there is a need for workers to retain the skills and knowledge to perform the task manually (Arora & Garg, 2024; Casner, Geven, Recker & Schooler, 2014; Volz & Dorneich, 2020).

*Managing workload*

An individual's workload is based on both their task load (how much work they need to achieve) and their capacity in the moment to achieve it (Gopher & Donchin, 1986). Reducing workload is one of the primary motivations for automating processes and evidence suggests this is the case when applied to automated information acquisition or analysis and to routine decisions or actions (Balfe, Sharples & Wilson, 2015; Endsley, 2017; Onnasch et al., 2014; Schreckenghost, Holden, Greene, Milam & Hamblin, 2023; Tatasciore et al., 2020). For non-routine or complex tasks, where there is a greater risk that the AI will fail and the human will need to intercede, full automation may do little to reduce overall workload, or it may introduce workload spikes and troughs (Bainbridge, 1983). There is some evidence that lower levels of automation are preferred when workload is high and the task is complex (Gutman, Olatunji & Edan, 2021).

*High stakes and social or ethical imperatives*

Parasuraman et al. (2000) recommended low to moderate levels of automation for decision and action implementation in high stakes scenarios, citing the potential negative consequences of out-of-the-loop errors if humans must take over from an incorrect or failed AI. People prefer to evaluate AI outputs under high risk scenarios (Loft et al., 2023). The use of automation has implications for how responsibility for a decision is attributed (Peng & Bhaskar, 2023; Shrestha, Ben-Menahem & von Krogh, 2019), and individuals should have control over the decisions for which they are held responsible, especially when stakes are high (Roth et al., 2019). High levels of automation have been associated with higher automation complacency (Cummings & Bruni, 2009), increasing the risk that incorrect automation decisions are missed. In other cases, social or ethical motivations may necessitate a need for direct human control in a task (Bigman & Gray, 2018). For example, automating the process of writing a manuscript, while technically feasible with a large language model, is prohibited by most publishers (Ganjavi et al., 2024).

*Meaningfulness and enjoyment*

Automating tasks that are dull or repetitive may have positive effects on a workers' overall sense of meaningfulness, by freeing up time for more significant tasks (Arora & Garg, 2024). However, a system that automates a task that workers find meaningful and enjoyable is unlikely to be met with enthusiasm. Automation can also reduce key factors contributing to meaningfulness, such as autonomy, skill variety and task significance (Arora & Garg, 2024). Danaher and Nyholm (2021) outline the impact of automation on workers' sense of achievement, noting that full automation is all but guaranteed to reduce achievement associated with the automated task. Although partial automation is also likely to impact achievement, it will depend on the worker's remaining contribution to the task and, especially, their autonomy (Scripter, 2024).

**Existing challenges and support needs**

Finally, decisions around automation mode may vary depending on the reason to consider AI in the first place: the challenges currently faced and the needs of the stakeholders. We categorised four broad categories.

*Need for scale*

The need to repeat a task at a much greater scale is perhaps the strongest motivation for employing automation. Human information processing constraints place limits on the maximum scale that can be achieved without some form of automation. Although some hybrid options are possible – for example, the AI and human split the load, the human sets rules to follow in the case of different task instances, or the human makes decisions over a batch of task instances – all require that some instances of the task will be run without human intervention.

*Need for efficiency*

Need for efficiency refers here to a need to perform the same amount of work in less time or with fewer resources. In ideal cases, full automation will typically be very efficient. However, this needs to be balanced against the risk that the automation fails and a human must correct the error, which may incur significant time/resource costs (Endsley & Kaber, 1999). Augmentation may improve efficiency depending on the nature of the task and source of the delays: for example, automating information acquisition and analysis may speed up the human decision loop (Tatasciore et al., 2020). Compared with augmentation and automation, collaboration may introduce new task components (i.e. additional back-and-forth interactions) that increase task resources.

*Need to maintain accuracy*

Third is the need to maintain a high standard of accuracy or quality of the task. If accuracy is important on a task, individuals are often more cautious in accepting AI advice (Eisbach, Langer & Hertel, 2023; Gregor & Benbasat, 1999). This suggest that, assuming humans are capably achieving high accuracy manually, their preference will be to remain in the loop. When quality is less important, relying on automated systems to enable efficiency or scale may be a suitable trade-off. We focused on the maintenance of accuracy rather than the need to *improve* accuracy because the latter will likely depend on AI capabilities, outside of the current scope. However, some hypothesise that when humans and AI are accurate in different ways (i.e. make different kinds of errors), overall accuracy may be improved by combining efforts through augmentation or collaboration (Choudhary et al., 2023).

*Need for innovation*

Finally, some tasks may involve what we call a need for innovation: the need to create something new, such as a product or work of art, or to make a new discovery. Although there have been some claims of using AI to fully automate breakthroughs (e.g., in science, Lu et al., 2024), these examples are limited. Researchers suggest innovation benefits from humans working with AI, bringing their curiosity, contextual understanding and discernment to recognise genuine innovation (Alesadi & Baiz, 2023; Bilgram & Laarmann, 2023; Leslie, 2023). Here the interaction is often couched as a collaboration, exploiting interactive capabilities of AI (generative AI in particular) to enable iterative steps towards a final product or discovery. Indeed, software tools aiming to support collaborative creation of products and content are showing promising results (e.g. Heyman et al., 2024).

**Analysis: Evaluating case studies against criteria**

The goal of the second phase of this study was to convert the criteria identified in the literature review into a rubric for recommending A$^2$C mode, and trial the rubric on sample tasks. Applying the rubrics require a decent level of knowledge about the task and the expertise of workers. To this end we used outputs from Cognitive Task Analysis (CTA), which, in addition to identifying the components and flow of a task, places emphasis on the cognitive demands faced by the worker and the accompanying strategies for solving problems and making decisions (Militello & Hutton, 1998). The emphasis on cognitive work makes CTA especially valuable for knowledge work domains or those with high levels of expertise (Gordon & Gill, 1997).

**A$^2$C criteria development and review**

Based on the literature review, we compiled a set of criteria and added clarifying questions that an analyst may ask when evaluating a task. We attempted to frame questions to elicit a response that would be most informative for A$^2$C mode decisions. For example, in regard to task type, the literature suggests that automating decision tasks presents the most challenges to human performance, so we framed the clarifying questions as "Does this task involve a decision?", where a response of Yes would be evidence towards keeping a human in the loop (augmentation or collaboration). To support the goal of keeping the rubric reasonably simple, we limited questions to yes/no or low/medium/high responses. Table 1 shows the full set of criteria and responses.

As shown in the table, the majority of the criteria had implications for deciding when full automation mode may be suitable versus when a human should remain in the loop (i.e. augmentation or collaboration). Only a few were judged to be important for identifying when a collaborative interaction may be beneficial over augmentation. For some questions, a response was not associated with any clear recommendation. For example, Need for Scale was associated with a recommendation for automation, but there was no clear recommendation for any mode if scale was not required (therefore table only shows outcomes for a "Yes" response).

To compile recommendations across the criteria, we trialled a simple points system, where responses were scored 0 points if automation was suitable (or most beneficial), 2 if human-in-the-loop (either augmentation or collaboration) was recommended and 4 if collaboration was recommended. While this approach implies a unidirectional and linear relationship between automation, augmentation and collaboration, which does not accurately capture the relationships between modes, we trialled this approach to investigate whether it may at least provide ballpark estimates of which mode was supported by the bulk of scores. We also recognised that some criteria held stronger implications for A$^2$C mode than others (e.g. the presence of high stakes was judged to be of greater implication than task complexity), and there Individual scores for each criterion were weighted based on its judged importance and scores were summed (weights are shown in Table 2).

**Case study methods and analysis**

Our data for evaluating the A$^2$C criteria were CTA outputs from three case studies for three science case studies. Details about the case studies are discussed briefly in the following section and a more detailed description of the CTA results of the genome annotation and protein crystallisation case

**Table 1**

*A²C scoring rubric*

| Criterion | Scoring questions | Response options and recommendations | | |
|---|---|---|---|---|
| | | **Automation is suitable or recommended** | **Augmentation or collaboration is recommended** | **Collaboration is recommended** |
| **Elements of the task** | | | | |
| Decision task | *Does this task involve a decision?* | No | Yes | |
| Component complexity | *How complex is this task in terms of 1) the number of different pieces of information that must be considered, 2) the number of steps in the task/actions that are taken?* | Low (few steps, info) | Medium or High (med-high steps, info) | |
| Dynamic complexity | *How dynamic is the state of the world in which the task takes place? How much impact do these have on completing the task?* | Low (static) | Medium or High (unpredictable) | |
| Coordinative complexity and interdependence | *How complex is the coordination/scheduling of the task? How interdependent are task components (i.e. does changing one affect how others are conducted)?* | Low (independent, linear flow) | Medium (some dependencies) | High (complex dependencies) |
| Variability | *How much do instances of this task vary from each other? Consider 1) variation in the problem (i.e. different starting conditions or information to consider) and 2) variation in the actions performed.* | Low (routine) | Medium (non-routine) | High (non-routine & novel) |
| Uncertainty: lacking information | *How much uncertainty is experienced due to a lack of information? (e.g. is the ground truth is available? Is there enough information to get a clear picture of the problem?)* | Low | Medium (manageable) | High (un-manageable) |
| Uncertainty: lacking understanding | *How much uncertainty is experienced due to a lack of understanding? (e.g. is it clear what action is best for a given instance? Are the underlying rules of cause/effect known?)* | Low | Medium (manageable) | High (un-manageable) |
| Presence of non-codified knowledge | *Does this task require knowledge that is not easily codified (e.g. experience, common sense, intuition, perceptual judgements)?* | No | Yes | |

**Impact on workers**

| | | | |
|---|---|---|---|
| Maintaining situation awareness | *Is an awareness or knowledge of what happens in this task required (e.g. for subsequent tasks or to check automated output)?* | No | Yes |
| Maintaining skills | *Are the skills developed through this task used on other occasions (e.g. to perform the manually on occasion or step in to fix an automated error)?* | No | Yes |
| Managing workload | *Are workers currently experiencing workload that is too high or unmanageable (i.e. more work than workers feel they have the capacity to complete)?* | Yes | |
| Risks | *Is this considered a high stakes task with serious consequences if something goes wrong?* | No | Yes |
| Social/ethical imperatives | *Are there social or ethical reasons to prioritise human decision-making in this task context?* | No | Yes |
| Motivation and enjoyment | *Do the people performing this task find significant value or meaning in manually performing this task?* | No | Yes |

**Challenges and support needs**

| | | |
|---|---|---|
| Need for scale | *Is there a need to perform the task at a significantly greater scale?* | Yes |

| | | | | |
|---|---|---|---|---|
| Need for efficiency | *Is there a need to reduce the amount of time or resources spent on this task?* | Yes | | |
| Need for maintaining accuracy | *Is there a need to maintain the current standard of accuracy, precision or quality?* | No | Yes | |
| Need for innovation | *Is there a need to create a new product or make a breakthrough or novel discovery in this task?* | | | Yes |

studies can be found in [anon. author reference]. In brief, data collection was based on interviews conducted with domain experts (protein crystallisation: $N$ = 4 experts in molecular biology; genome annotation: $N$ = 12 genomics and bioinformatics researchers; biological collections: $N$ = 8 biological curators or digital curators, all working within [anon. organisation]). The genome annotation and protein crystallisation use cases followed a variant of the Applied Cognitive Task Analysis methodology (Militello & Hutton, 1998) where participants completed individual interviews to work process and the cognitive demands they faced, following by a deepening phase to identify more information about their decision-making process (based on the Critical Decision Method; Klein, Calderwood & MacGregor, 1989). For the biological collections use case, due to participant time constraints, data was collected through a group interview and observation of the task with all participants. All interviews focused on identifying 1) participants' current workflow, 2) challenges faced and 3) information, knowledge and strategies applied to mitigate challenges. These data were compiled in cognitive demands tables (Militello & Hutton, 1998) for each participant and then combined across participants (except biological collections, in which one a single table were generated for the group).

We noted that the workflow of each case study was comprised of three distinct steps. Each step was considered a discrete task for the purpose of A$^2$C scoring (see the following section for steps).

To score the nine tasks (three tasks per case study), three researchers first scored one candidate task on the criteria. Researchers drew on the CTA outputs (cognitive demands tables, flow diagrams and notes) to arrive at scores. This was performed individually then evaluated as a group. Group discussions led to adjustments in the wording and relative weighting of different factors. We then conducted an analysis of all tasks. Each task was reviewed by two researchers, who then met to resolve any conflicting scores. Where scores were polar opposites (e.g. one researcher scored yes and the other no, or one scored low and the other high), the researchers discussed their reasons for the score and agreed on an appropriate response. This accounted for approximately 20% of scores. Where one researcher selected medium and the other low or high (another 15% of scores), the differing scores were kept and simply averaged in the summation process.

**Case studies and scores**

We now briefly summarise the case studies alongside the scores for each task (see Table 2). Viewed across all case studies, we found that scores fell into three clusters: one task with a much lower score (16) compared with all other tasks, five intermediate tasks (scoring 24-30) that could be considered candidates for augmentation, and three tasks each with a higher score of 35 that emerged as the best candidates for collaboration.

**Table 2**

*Use case tasks score on the A²C rubric*

| Legend | Automation is suitable or recommended | Augmentation or collaboration is recommended | Collaboration is recommended |

| | | Case study | Protein crystallisation | | | Genome annotation | | | Digitisation of biological collections | | |
|---|---|---|---|---|---|---|---|---|---|---|---|
| **Criterion** | **Score weight** | | Select initial screen | Determine if crystal formed | Select optimising screen | Structural annotate | Functional annotate | Data quality control | Image specimens | Transcribe metadata | Metadata quality control |
| Decision task | 1 | | Y | Y | Y | Y | Y | Y | N | N | Y |
| Component complexity | 1 | | M | L | M-H | M | M | M-H | M | L-M | M |
| Dynamic complexity | 1 | | L | H | M-H | L-M | L-M | L-M | L | L | M |
| Coordinative complexity | 1 | | M-H | M-H | M-H | L | L-M | M-H | M | L | L-M |
| Variability | 1 | | M-H | L | M-H | M | M | M | L | M | M |
| Uncertainty: information | 1 | | L-M | L-M | M-H | M | M-H | M | L | L-M | L-M |
| Uncertainty: understanding | 1 | | M-H | L | H | L | L-M | M | L | L | L |
| Non-codified knowledge | 1 | | Y | Y | Y | Y | Y | Y | N | N | Y |
| Situation awareness | 1 | | N | N | N | Y | Y | Y | N | Y | N |
| Maintaining skills | 1 | | N | N | N | Y | Y | Y | N | N | N |
| Managing workload | 3 | | N | N | N | Y | Y | Y | Y | Y | Y |
| Risks | 3 | | N | N | N | N | N | N | Y | Y | Y |
| Social/ethical imperatives | 2 | | N | N | N | N | N | N | N | N | N |
| Motivation and enjoyment | 2 | | N | N | N | Y | Y | N | N | Y | N |
| Need for scale | 1 | | N | N | N | Y | Y | Y | Y | Y | Y |
| Need for efficiency | 1 | | Y | Y | Y | N | Y | N | Y | Y | N |
| Need for accuracy | 1 | | Y | Y | Y | Y | Y | Y | Y | Y | Y |
| Need for innovation | 1 | | N | N | N | Y | Y | N | N | N | N |
| **Total score** | | | 30 | 24 | 35 | 35 | 35 | 30 | 16 | 25 | 29 |
| **A²C cluster** | | | Aug | Aug | Collab | Collab | Collab | Aug | Auto | Aug | Aug |

*Note.* Colour coding of cells corresponding with the point score assigned to the value, from 0 points (lightest grey) through to 4 points (darkest grey).

*Protein crystallisation*

Protein crystallisation involves subjecting samples of protein to different experimental conditions with the aim of causing the protein to crystallise. The crystals can then be passed through X-ray diffraction to identify protein structure. Achieving crystals is challenging because the specific experimental conditions that produce crystals cannot be predicted in advance, and it typically require trial and error with a wide range of different experimental factors. We identified three key tasks within protein crystallisation:

1) Selecting the initial set of conditions (or "screen")
2) Identifying whether crystals have formed in each experiments, and whether those crystals are good enough to be sent to X-ray diffraction
3) If promising crystals are identified, but are not yet good enough to X-ray, select follow-up "optimisation" screens to try to achieve higher quality crystals

Scoring with the rubric (see Table 2) indicated the final stage, selecting an optimisation screen, was associated with the highest score, equal highest score across the three case studies, suggesting it was the best candidate for collaboration. Selecting optimisation screens is associated with high levels of uncertainty due to a lack of knowledge about the causal relationships between experimental factors and outcomes (lack of understanding). The first and second stages, tasks that are somewhat less complex, were candidates for some AI support through augmentation.

*Genome annotation*

Genome annotation is the process of identifying the structure and function of gene sequences and is often conducted by genomics or bioinformatics researchers as part of their work. Analysis of the genome annotation process revealed significant variation amongst researchers in their workflow, however three broad tasks emerged:

1) Identifying individual genes within sequences, typically through a combination of automated detection algorithms and biological knowledge of gene components.
2) Identifying the function of gene structures, typically by searching function databases for any closely matching genes and adopting the same function.
3) Performing quality control over the final dataset, particularly those parts of the process that have been automated.

Scoring revealed that the three stages had relatively similar, high scores, suggesting little role for full automation in genome annotation. Structural and functional annotation were considered candidates for collaboration due largely to high uncertainty and the need to make new discoveries. For example, in functional allocation, when the database lookup approach fails, researchers expressed a desire to be able to seek out new functional information through other sources, such as literature.

*Digitising Biological Collections*

Many large collections of biological specimens are currently seeking to digitise their collection and enter metadata (e.g. specimen species, date of collection) from the physical label on the specimen into a database. Interviews with curators revealed three primary steps in the digitisation process:

1) Imaging the specimen and the labels
2) Transcribe the metadata from the label image into a database (a process often performed by citizen scientists)
3) Performing quality control on the transcribed data (typically performed by digital curators)

The initial imaging task emerged as a good candidate for automation, with the lowest score across the case studies. This is a relatively routine task involving positioning and photograph specimens and adjusting digital photo files. Both metadata transcription and quality control scores suggested input from a human was required, due largely to the increased cognitive complexity, though there was little evidence that collaboration would be beneficial for either of these tasks.

## Discussion

The rapid evolution and dissemination of AI tools for knowledge work is leaving many in the industry struggling to integrate new AI tools into existing workflows. Human factors researchers have been considering the implications of AI teaming for many years and offer a wealth of expertise. However, simple, practical guidance for those outside the domain is difficult to find, particularly when making decisions regarding the mode of interaction appropriate for a given task.

The current study had three main research questions, and we discuss each in turn below.

**RQ1: What are the criteria to guide selection of human-AI interaction mode?**

Our first goal was to compile a set of key criteria for human-AI interaction mode, conceptualised using the A$^2$C framework of automation, augmentation and collaboration. We developed an list of 18 factors, drawing on literature from human factors, computer science and information sciences. We see this as a preliminary and living list, as with technology evolution and continuing research these recommendations will likely develop. Currently there is a body of research on function allocation and the appropriate levels of automation (per the Parasuraman et al. 2000 framework, for example), which provided good guidance for the decision between full automation and human-in-the-loop.

However, we found less guidance regarding when collaboration should be used over augmentation. Unlike the distinction between automation and human-in-the-loop, the augmentation vs collaboration distinction comes down to the nature of the interaction. Collaboration involves reciprocal information sharing and the capacity to perform different functions as required to move towards a common goal. More evidence on the benefits or risks of using software to support collaboration is required.

Building on the work of Tariq et al. (2024), we hypothesised four factors that may be discriminative in deciding when a collaborative interaction may be most appropriate, pending further research.

1) *The task is likely to involve novel instances*. Novel task instances may require additional data collection or analysis, though it may not be possible to determine which information or actions will be most required in advance. Therefore, an AI system that conduct different information seeking tasks as needed, while keeping track of progress towards the goal, may be most effective. Through a collaborative interaction, the user can exploit their more sensitive awareness of the situation context to guide the direction of the interaction and decide when a solution to the problem has been realised.

2) *Decision uncertainty is at unacceptable levels.* Augmentation can help reduce uncertainty by collecting information and making recommendations to present to the user. However, if uncertainty remains high, a common first strategy is to seek more information (Lipshitz & Strauss, 1997). What is needed in this case is not to cycle through the same augmentation process, but to seek new information that complements the existing knowledge. As with novel tasks, a collaborative solution keeps the user in the driver's seat, directing information and problem-solving activities.

3) *Coordinative complexity or interdependence is high.* When tasks are highly interdependent and have complexity coordination needs, it may be impractical to disentangle discrete components to assign to automation. A good collaborative system should be capable of handling multiple connected components and flexible enough to adapt to changes in one part of the process. Human-AI collaboration often involves some form of interdependence; for example, each agent relies on complementary input from another agent to complete its action (Verhagen, Marcu, Neerincx & Tielman, 2024). Indeed, recent evidence suggests that people prefer working interdependently with AI more than working independently with AI (Li, Huang, Liu, Zheng & 2022).

4) *There is a need for innovation.* Creating something new or making a new discovery involves working towards a novel outcome. The path towards innovation is unlikely to be linear or easily anticipated, requiring a flexible and multi-functional tool. Generative AI and its ability create content, suggest ideas and hone output in response to user input may be especially value.

**RQ2: Can we evaluate a given task against these criteria using task analysis?**

Our second goal was to evaluate whether a given task could be scored against these criteria using a relatively efficient knowledge elicitation approach. We found that applied cognitive task analysis offered good insight into the nature of the task, including elements of complexity and variability. This method targets challenging elements of tasks, which enabled us to identify support needs, and the focus on cognitive expertise provided high levels of detail regarding uncertainty, situation awareness needs and reliance on non-codified knowledge. Areas our data collection did not specifically target included risks and social or ethical issues around automation. Nonetheless, enough information emerged naturally in interviews that we felt we were able to answer these. A potential limitation with cognitive task analysis is that it is structured to expressly focus on decision points in the workflow or tasks that a cognitively demanding. Using only this information may miss opportunities for automation of less cognitively demanding, but nevertheless challenging, tasks.

We expect that a range of different methods could be used to elicit enough information to address the rubric. For small-scale or rapid projects that are unlikely to have the resources for thorough analysis, the criteria could provide a simple protocol for non-human factors experts to run a knowledge elicitation session with subject matter experts.

**RQ3: (How) can we then bring together these criteria to make a recommendation about interaction mode needed?**

The goal of the $A^2C$ rubric is to support non-human factors specialists to make AI design decisions for a specific task. We trialled a system to score each of the responses to the criteria and to sum the responses to support a final recommendation. The scoring policy required some assumptions about the relative weight of different factors, and the validity in using a summative approach. Future work will test and iterate on this approach, and we see these as parameters that practitioners can adapt as required. Inspection of the results suggests the scoring was broadly capturing the expected findings: the tasks with the lowest scores were routine and of lower complexity while the highest scores were on tasks associated with high complexity and uncertainty.

A number of elements complicate the recommendation process. One is that factors can interact. For example, situation awareness needs interact with task type (Endsley, 2017). Incorporating interactive effects is feasible but would require a more complex scoring formula than currently implemented. Additionally, some factors may have overriding importance: if there is a social or ethical reason to keep the human in the loop, full automation may never be an option, regardless of how other factors emerge. Moreover, there is additional nuance that is intentionally ignored in a simple rubric. For example, in Need for Efficiency, we noted that automation is likely more beneficial than augmentation. In practice this likely depends on what part of the task is causing a bottleneck: if it stems from inefficiency information acquisition or analysis rather than decision process, augmentation may be just as beneficial as automation.

Ultimately, a prescriptive use of this collection of criteria may be too simplistic. Its value may lie in signposting critical pieces of information that developers should explore and providing a place to bring this information together to guide design decisions. For many system developers, the first instinct may be to build full automation solutions to problems, and ideally this approach may lead developers to pause and rethink alternative options.

**Limitations and next steps**

This study represents the initial step toward developing a tool intended for practical use, and more work is needed to evaluate its validity and gain input from subject matter experts. While we focus here on the domain of science, we see this approach as broadly applicable across knowledge work domains and seek to validate the findings across different domains. Future iterations will also reexamine the criteria list; for instance, an additional criterion under consideration concerns how errors are handled in the task. When errors necessitate timely human takeover, it becomes much more critical that humans develop situation awareness and maintain skills, and therefore this factor may be worth explicitly considering. Finally, as mentioned earlier, future iterations could be expanded to encompass factors related to the suitability of different technology options, which have been incorporated in other AI use case surveys (Engel et al., 2024).

23
AI IN WORKFLOWSEndsley, M. R. (1995). Toward a Theory of Situation Awareness in Dynamic Systems. *Human Factors: The Journal of the Human Factors and Ergonomics Society*, *37*(1), 32–64. https://doi.org/10.1518/001872095779049543

Endsley, M. R. (1997). Level of Automation: Integrating Humans and Automated Systems. *Proceedings of the Human Factors and Ergonomics Society Annual Meeting*, *41*(1), 200–204. https://doi.org/10.1177/107118139704100146

Endsley, M. R. (2017). From here to autonomy: Lessons learned from human–automation research. *Human Factors*, *59*(1), 5–27. https://doi.org/10.1177/0018720816681350

Endsley, M. R., & Kaber, D. B. (1999). Level of automation effects on performance, situation awareness and workload in a dynamic control task. *Ergonomics*, *42*(3), 462–492. https://doi.org/10.1080/001401399185595

Endsley, M. R., & Kiris, E. O. (1995). The out-of-the-loop performance problem and level of control in automation. *Human Factors*, *37*(2), 381-394. https://doi.org/10.1518/00187209577906455

Engel, C., Elshan, E., Ebel, P., & Leimeister, J. M. (2024). Stairway to heaven or highway to hell: A model for assessing cognitive automation use cases. *Journal of Information Technology*, *39*(1), 94–122. https://doi.org/10.1177/02683962231185599

Frohm, J., Lindström, V., Stahre, J., & Winroth, M. (2008). Levels of automation in manufacturing. *Ergonomia-an International Journal of Ergonomics and Human Factors*, *30*(3). https://www.diva-portal.org/smash/record.jsf?pid=diva2:216233

Ganjavi, C., Eppler, M. B., Pekcan, A., Biedermann, B., Abreu, A., Collins, G. S.,… & Cacciamani, G. E. (2024). Publishers' and journals' instructions to authors on use of generative artificial intelligence in academic and scientific publishing: Bibliometric analysis. *BMJ*, *384*, e077192. https://doi.org/10.1136/bmj-2023-077192

Goh, Y. M., Micheler, S., Sanchez-Salas, A., Case, K., Bumblauskas, D., & Monfared, R. (2020). A variability taxonomy to support automation decision-making for manufacturing processes. *Production Planning & Control*, *31*(5), 383–399. https://doi.org/10.1080/09537287.2019.1639840

Gopher, D., & Donchin, E. (1986). Workload: An examination of the concept. In *Handbook of perception and human performance, Vol. 2: Cognitive processes and performance* (pp. 1–49). John Wiley & Sons.

Gordon, S. E., & Gill, R. T. (2014). Cognitive task analysis. In *Naturalistic Decision Making* (pp. 131-140). Psychology Press.

Gregor, S., & Benbasat, I. (1999). Explanations from intelligent systems: Theoretical foundations and implications for practice. *MIS Quarterly*, 497–530. https://doi.org/10.2307/249487